\documentclass[twocolumn,times,trackchanges]{aastex631}
\hypersetup{linkcolor=purple,citecolor=cyan,filecolor=cyan,urlcolor=blue,hyperfigures}

\newcommand{\speed}[1]{#1 km~s${}^{-1}$}
\newcommand{\nfig}[1]{Figure~\ref{#1}}
\shorttitle{Coronal Plasma Flows during Stellar Flares of EV Lac}
\shortauthors{Chen et al.}
\watermark{DRAFT}
\usepackage{CJK}

\begin{document}

\title{Detection of Flare-induced Plasma Flows in the Corona of EV Lac with X-ray Spectroscopy}
\begin{CJK*}{UTF8}{gbsn}
\correspondingauthor{Hui Tian}
\email{huitian@pku.edu.cn}

\author[0000-0001-7866-4358]{Hechao Chen (陈何超)}
\affil{School of Earth and Space Sciences, Peking University, Beijing, 100871, China.}

\author[0000-0002-1369-1758]{Hui Tian (田晖)}
\affil{School of Earth and Space Sciences, Peking University, Beijing, 100871, China.}

\author[0000-0001-5612-4457]{Hao Li (李昊)}
\affil{Instituto de Astrof\'{\i}sica de Canarias, E-38205 La Laguna, Tenerife, Spain}
\affil{Departamento de Astrof\'{\i}sica, Universidad de La Laguna, E-38206 La Laguna, Tenerife, Spain}

\author[0000-0003-4156-3793]{Jianguo Wang (王建国)}
\affil{Yunnan Observatories,Chinese Academy of Sciences, 396 Yangfangwang, Guandu District, Kunming, 650216, China}

\author{Hongpeng Lu (陆洪鹏)}
\affil{School of Earth and Space Sciences, Peking University, Beijing, 100871, China.}

\author[0000-0002-7421-4701]{Yu Xu (徐昱)}
\affil{School of Earth and Space Sciences, Peking University, Beijing, 100871, China.}

\author[0000-0003-4804-5673]{Zhenyong Hou (侯振永)}
\affil{School of Earth and Space Sciences, Peking University, Beijing, 100871, China.}
     
\author[0000-0001-8487-7587]{Yuchuan Wu (吴昱川)}
\affil{School of Earth and Space Sciences, Peking University, Beijing, 100871, China.}



\begin{abstract} 
Stellar flares are characterized by sudden enhancement of electromagnetic radiation from the atmospheres of stars. Compared to their solar counterparts, our knowledge on the coronal plasma dynamics of stellar flares and their connection to coronal mass ejections (CMEs) remains very limited. With time-resolved high-resolution spectroscopic observations from the \textit{Chandra} X-ray observatory, we detected noticeable coronal plasma flows during several stellar flares on a nearby dMe star EV Lac. In the observed spectra of O~{\sc{viii}} (3 MK), Fe~{\sc{xvii}} (6 MK), Mg~{\sc{xii}} (10 MK), and Si~{\sc{xiv}} (16 MK) lines, these flare-induced upflows/downflows appear as significant Doppler shifts of several tens to \speed{130}  {, and the upflow velocity generally increases with temperature}. Variable line ratios of the Si~{\sc{xiii}} triplet reveal that these plasma flows in most flares are accompanied by an increase of the coronal plasma density and temperature. We interpret these results as X-ray evidences for chromospheric evaporation on EV Lac. In two successive flares, the plasma flow pattern  {and a sharp increase of the measured coronal density are highly suggestive of explosive evaporation.} The transition from redshifts to blueshifts in such an explosive evaporation occurs at a temperature of at least 10 MK, much higher than that observed in solar flares ($\sim$1 MK). However, in one flare the cool and warm upflows appear to be accompanied by a decreasing plasma density, which might be explained by a stellar filament/prominence eruption coupled to this flare. These results provide important clues to understand the coronal plasma dynamics during flares on M dwarfs.

\end{abstract}



\section{Introduction} \label{sec:intro}
In the standard model of solar flares \citep[e.g.,][]{2010ARA&A..48..241B,2011SSRv..159...19F,2011LRSP....8....6S}, a considerable amount of magnetic energy is rapidly released through magnetic reconnection in the form of accelerated electrons. These energized electrons travel downward along closed magnetic loops into the lower atmosphere, causing intense chromosphere heating. The ensuing overpressure explosively drives hot evaporating flows with millions of Kelvin into the corona. 
Such an ongoing ``chromospheric evaporation" process \citep{1985ApJ...289..425F} then results in a delayed soft X-ray (SXR) and extreme ultraviolet (EUV) flare emission in post-flare loops \citep{1968ApJ...153L..59N,1993SoPh..146..177D}. Above the post-flare loops, an ejecting plasmoid, i.e., filament/prominence, sometimes may appear and lead to a coronal mass ejection (CME) after a rapid acceleration \citep[see][]{1999Ap&SS.264..129S,2000JGR...10523153F,2000JGR...105.2375L,2003NewAR..47...53L}. 
Due to the similarity between stellar and solar flares that revealed from previous photometric observations \citep[e.g.,][]{1968ApJ...153L..59N,1993SoPh..146..177D,1995ApJ...453..464H,1996ApJ...471.1002G} and recent parameter analyses \citep{2013ApJ...771..127N,2013ApJS..209....5S,2017PASJ...69...41M,2017ApJ...851...91N,2021MNRAS.505L..79Y}, stellar flares are generally assumed to be caused by the same physical processes as described in the standard model of solar flares. However, this assumption has not been tested in detail yet.

As the key supportive evidence for the standard model of solar flares, chromospheric evaporations during solar flares have been well studied through spatially-resolved spectral observations at optical, ultraviolet (UV), EUV and SXR wavelengths \citep[e.g.,][]{2003ApJ...586.1417B,2003ApJ...588..596T,2015ApJ...807L..22G,2015ApJ...811....7L,2015ApJ...811..139T,2016ApJ...827...27Z,2017ApJ...841L...9L,2017ApJ...848..118L,2018ApJ...854...26L,2018ApJ...856...34T,2018PASJ...70..100T,2020ApJ...896..154Y,2021ApJ...921L..33Y}. According to the energy flux injection rate, chromospheric evaporations are divided into explosive and gentle ones  \citep{1985ApJ...289..425F}. Explosive evaporations are characterized by upflows with speeds up to several hundred and are associated with obvious cool downflows called ``chromospheric condensation" \citep{1984SoPh...93..105I,1990ApJ...348..333C,1993ApJ...416..886G,2019ApJ...879...19Z}. On the contrary, gentle ones only show upflows with speeds of several \speed{tens} and do not reveal any associated downflows. For explosive ones, a reversal of blue/redshifts is often found in the temperature range of 0.8$-$2 MK  on the Sun \citep[e.g.,][]{2006ApJ...638L.117M,2008ApJ...680L.157M,2010ApJ...724..640C,2015PhPl...22j1206I,2019ApJ...875...93C}. This transition temperature sometimes even reaches up to 5 MK \citep{2011ApJ...727...98L}. Moreover, the velocity of an evaporating flow normally increases with the spectral line formation temperature, which can reach up to \speed{200$-$400} at temperatures of more than 10 MK \citep[e.g.,][]{1982SoPh...78..107A,2014ApJ...797L..14T,2015ApJ...813...59L,2015ApJ...799..218Y,2015ApJ...811..139T,2019ApJ...870..109Z}.  
By comparison, on other stars, hot evaporation flows during flares have been rarely detected, as reviewed in \citep{2009A&ARv..17..309G,2010SSRv..157..211G,2019LNP...955.....L}.
Indirect ways such as coronal density increase indicated by He-like triplet analysis \citep[e.g.,][]{2002ApJ...580L..73G,2010A&A...514A..94L}, ``Neupert effect" inferred from multi-band light curves \citep[e.g.,][]{1996ApJ...471.1002G,1995ApJ...453..464H,2003ApJ...597..535H}, and abundance changes \citep[e.g.,][]{2009ApJ...707L..60L,2010A&A...514A..94L} were used to infer the existence of such plasma flows. 

On the Sun, flares are often accompanied by CMEs \citep[see][]{2011LRSP....8....1C}. Solar CMEs are known as the main driver of severe space weather disturbances \citep{1993JGR....9818937G} and have been routinely imaged by white-light coronagraphs \citep{2012LRSP....9....3W}. 
The possibility of CME detection through Sun-as-a-star observations has also been demonstrated \citep{2016ApJ...830...20M,2022arXiv220411722X,2022arXiv220403683Y}.
As their counterparts, stellar CMEs have also gained more and more attention in recent years, because of their potential hazard to the exoplanetary space weather \citep{2020IJAsB..19..136A}. 
But so far only a handful of possible stellar CMEs have been detected through several methods \citep[see details in][]{2019ApJ...877..105M}, including significant plasma motions revealed by time-resolved spectroscopy \citep[e.g.,][]{1990A&A...238..249H,2016A&A...590A..11V,2019NatAs...3..742A,2021NatAs.tmp..246N}, X-ray dimmings after stellar flares \citep{2021NatAs...5..697V}, as well as X-ray continuum absorption \citep{2017ApJ...850..191M}. Several numerical simulations have attempted to study the coronal/interplanetary responses of stellar CMEs and their potential impacts on nearby exoplanets \citep[e.g.,][]{2019ApJ...884L..13A,2019ApJ...880...97L,2020IAUS..354..426J,2022ApJ...924..115O}. The possible strong magnetic suppression and the observed low CME occurrence rates on M dwarfs have also been discussed  by several authors \citep{2016IAUS..320..196D,2018ApJ...862...93A,2021ApJ...917L..29L,2022MNRAS.509.5075S}.

In order to understand the plasma dynamics during stellar flares and their connection to CMEs, many spectroscopic studies of stellar flares have been conducted in optical and UV passbands. In these studies, apart from line broadenings \citep[e.g.,][]{2004A&A...420.1079F,2013A&A...560A..69L,2020PASJ...72...68N,2021ApJ...916...92W,2022ApJ...928..180W}, line asymmetries have also been frequently observed in low-temperature chromospheric (or transition-region) lines during stellar flares \citep[e.g.,][]{1993A&A...274..245H,1999A&A...349..863B,2004A&A...420.1079F,2008A&A...487..293F,2011A&A...536A..62L,2018PASJ...70...62H,2020A&A...637A..13M,2020MNRAS.499.5047M,2021A&A...646A..34K,2021PASJ...73...44M,2022ApJ...928..180W,2022arXiv220509972L}. For instance, \citet{1994A&A...285..489G} detected prominent evaporating chromospheric plasma flows in Balmer and Ca~{\sc{ii}} lines during a flare event on a dM4.5e star AT Mic. During and shortly after the impulsive phases of individual flares on AD Leo, \citet{2003ApJ...597..535H} detected signatures of chromospheric condensation in UV lines, with a Doppler velocity of \speed{40}.
To search for flares and CMEs on M dwarfs, \citet{2019A&A...623A..49V} and \citet{2020A&A...637A..13M} independently investigated a large number of stellar chromospheric spectra, but they consistently claimed a low detection rate of possible CMEs and a high level of flare activity on M dwarfs.
Recently, \citet{2021NatAs.tmp..246N} reported the first detection of an eruptive filament associated with a superflare (i.e., a flare with an energy of more than 10$^{33}$ erg) on EK Dra. The main signatures are a blueshifted absorption component with a high speed of \speed{510} in H$\alpha$ and a related dimming in the light curve of the H$\alpha$ equivalent width. Through a comparison between this eruption and solar filament eruptions, they concluded that this eruption very likely launches a stellar CME. 

Compared to spectral studies at optical/UV wavelengths, time-resolved SXR spectroscopic observations can better reveal the coronal physical processes behind stellar flares. However, only very few such observations exist \citep[see the review of][]{2004A&ARv..12...71G}.
The first X-ray spectral evidence for chromospheric evaporation during stellar flares was presented by \citet{2002ApJ...580L..73G}. This large flare was observed on Proxima Centauri with \textit{XMM-Newton}. During this flare, apart from a clear ``Neupert effect", they also found a strong density increase from a pre-flare level of n$_e<10^{10}$ to 4$\times 10^{11}$ cm$^{-3}$,  which was inferred from the He-like O~{\sc{vii}} triplet. Recently, using time-resolved SXR spectroscopic observations provided by \textit{Chandra} X-ray observatory, \citet{2019NatAs...3..742A} first unambiguously detected multi-temperature (about 3$-$25 MK) coronal plasma flows in a mega flare on HR 9024. They attributed the cool upflows detected in the O~{\sc{viii}} ( 18.97 \AA~) line to a possible CME coupled to the flare and the hotter upflows detected at the flare onset to chromospheric evaporation.

With time-resolved X-ray spectra of a nearby flare star EV Lac obtained with the \textit{Chandra} X-ray observatory, this study focuses on probing coronal plasma flows during flares, as well as their associated coronal density/temperature variations. We describe the instrument and observation in Section 2. The data analysis and methods are described in Section 3. Section 4 presents the results and interpretation in the context of the standard model of solar flares while Section 5 provides discussions and conclusion. 

\section{Instruments and Observation} \label{sec:obs}
\subsection{Instruments and Data Selection}
The \textit{Chandra X-ray Observatory} was launched on 1999-Jul-23 \citep{2000SPIE.4012....2W}. As one of the two high-resolution grating systems on the \textit{Chandra}, the High Energy Transmission Grating (HETG) consists of two sets of transmission gratings, i.e., the Medium Energy Grating (MEG) and the High Energy Grating (HEG), which simultaneously collect X-ray spectral signals in the wavelength ranges of 2.5$-$31 \AA~and 1.2$-$15 \AA, respectively. In conjunction with a spectroscopic array of the Advanced CCD Imaging Spectrometer (ACIS-S) detector \citep{2003SPIE.4851...28G}, the HETG forms the High-Energy Transmission Grating Spectrometer (HETGS) \citep{2005PASP..117.1144C}. The \textit{Chandra}/HETGS provides a very high spectral resolving power  ($\lambda$/$\Delta \lambda \sim$100$-$1000) and the high-accuracy wavelength calibration allows velocity measurements down to $\sim$\speed{10$-$20} \citep[see][]{2006ApJ...644L.117I,2017A&A...607A..14A}. 

The \textit{Chandra Transmission Grating Data Archive and Catalog} \citep[\textit{TGCat},][]{2011AJ....141..129H} provides easy access to observations of a particular object or type of object, and its web search interface also supports a quick review on the quality and potential scientific usefulness of the spectra products. In order to search for flare-induced coronal plasma flows and possible stellar CMEs, we checked the spectra products of strong X-ray flare events that occur on single stars with the aid of \textit{TGCat}, focusing on the \textit{Chandra} category ``Stars and WD". Finally, we found observations of two single stars, including HR 9024 and EV Lac, with obvious Doppler shifts of spectral lines. 
The former has been investigated by \citet{2019NatAs...3..742A}. Here, we investigated the plasma flow pattern during several flares on EV Lac and tried to explain their physical origins. 

\subsection{EV Lac and its \textit{Chandra} Observations}
EV Lac (EV Lacertae or Gl 873) is a nearby (5 pc) dM3.5e flare star in the constellation Lacerta, which emits strong X-rays and is known to frequently produce flares (with frequency up to 0.48 hr$^{-1}$) \citep[e.g.,][]{1984ApJ...284..270A,1997A&A...327.1114L,2005ApJ...621..398O,2021ApJ...922...31P,2020MNRAS.499.5047M}. The quiescent coronal temperature of EV Lac appears to be 3$-$20 MK, since its differential emission measure (DEM) distribution derived from the \textit{Extreme Ultraviolet Explorer} \citep[\textit{EUVE,}][]{1991AdSpR..11k.205B} and \textit{Chandra} observations peaks at log T/K $\sim$6.4 and  does not decrease too much till log T/K = 7.2 \citep{2006ApJ...647.1349O}.
Due to its rapid spinning ($\sim$4.3 days), EV Lac hosts stronger magnetic activities and can produce many more powerful flares (with energy up to 10$^{34}$ erg) compared to our Sun \citep{2000A&A...353..987F,2010ApJ...721..785O}. With time-resolved high-resolution H$\alpha$ spectroscopy of a flare on EV Lac, \citet{2018PASJ...70...62H} recently reported a blue wing enhancement during the whole duration of the flare (more than 1.5 hr) and an absorption component in the red wing during the early and later phases of the flare. They attributed the latter to chromospheric downflows in the post-flare loops, while tentatively ascribed the former to evaporation  or filament activation/activity. \citet{2020MNRAS.499.5047M} monitored EV Lac spectroscopically at a high resolution for 127 hours. They found a significant blue shift ($\sim$ \speed{220}) in one of 27 H$\alpha$ flares and ascribed it to an erupting filament.

EV Lac was observed by the \textit{Chandra}/HETGS twice in September 2001 for 100 ks (ObsID 1885) and March 2009 for 97 ks (ObsID 10679). Using these observations, \citet{2010ApJ...723.1558H} previously conducted a survey of all short- and long-duration flares on EV Lac, focusing on their photometric parameters (amplitude, shape, and scale), temperature, emission measure, and Fe K fluorescence. With these two observations, our current work aims to probe flare-associated plasma flows and plasma parameter variations. 
For brevity, we hereafter refer to these two observations as Obs1 and Obs2, respectively. We used the analysis-ready \textit{Chandra}/HETGS X-ray count spectra products obtained from the \textit{TGCat}.

\section{Data analysis and Methods}
\subsection{The X-ray light curves and composite X-ray spectra}
\begin{figure}
\centering
\includegraphics[width=1.\textwidth]{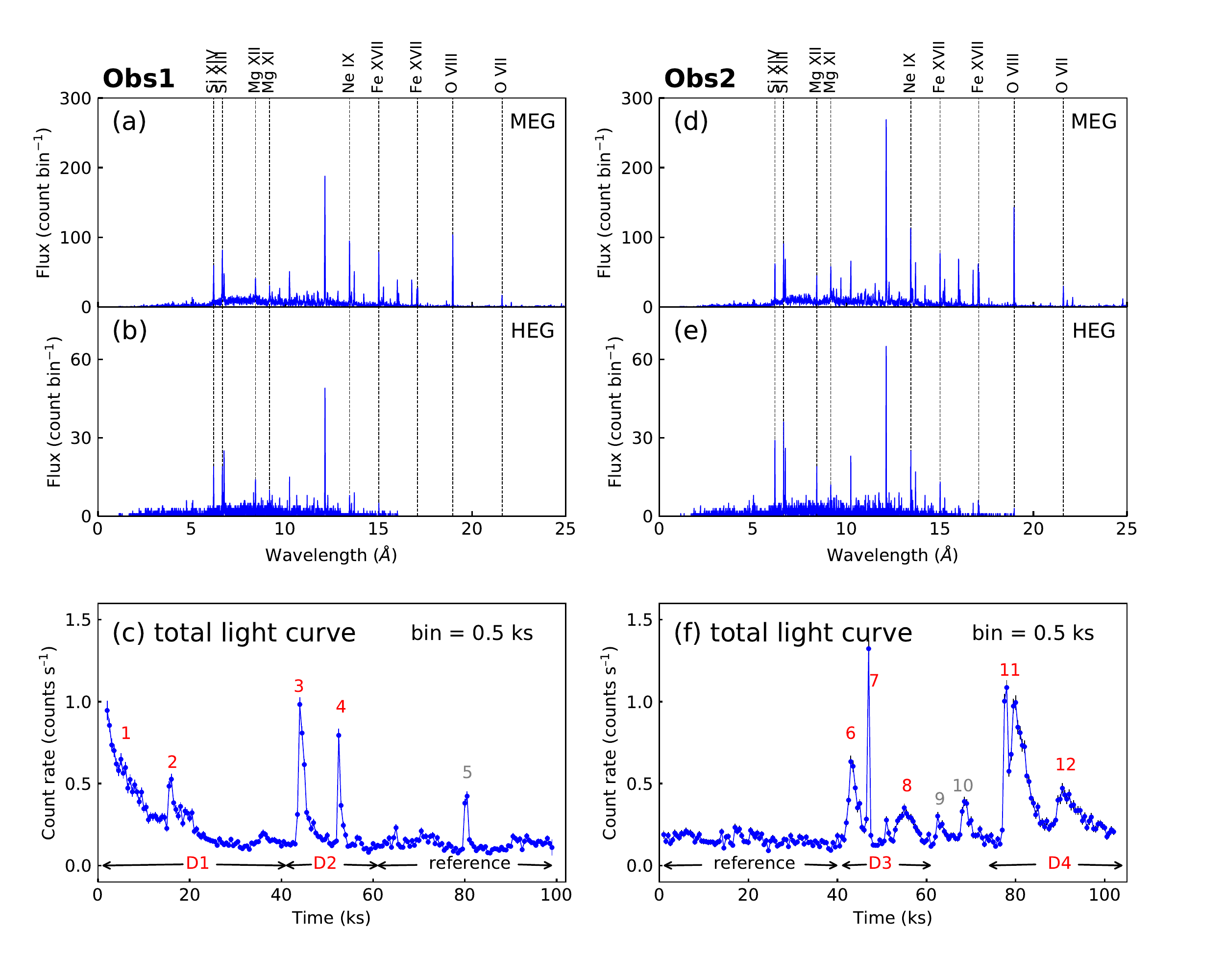}
\caption{Observed X-ray spectra and light curves of EV Lac.  (a) and (b) X-ray spectra integrated over the first Chandra observation (Obs1)  of the Medium Energy Grating (MEG) and High Energy Grating (HEG), respectively. The bin sizes of MEG and HEG are 5 m\AA~ and 2.5 m\AA, respectively. Spectral lines mentioned in the current work are marked.  (c) Light curve of X-ray flux summed over the wavelength range of (1$-$25 \AA) and the $\pm$1 diffraction orders of MEG and HEG. The time bin is 0.5 ks. (d-f) Similar to (a-c), but for the second observation (Obs2). ``D1", ``D2", ``D3", and ``D4" denote the time intervals with flare activity of our interest, while ``reference" denotes the time intervals without too much flare activity. Note that at least 12 flares can be easily identified in Obs1 and Obs2.
\label{fig1}}
\end{figure}

\nfig{fig1} presents the X-ray spectra and total light curves of EV Lac observed in Obs1 and Obs2. The total 200-ks-long observation of EV Lac includes at least 12 flares (as labeled by No.1$-$12), indicating a high level of flaring activity. In order to study the coronal dynamics of these flares with time-resolved spectroscopy, we extracted both HEG and MEG spectra as a function of time during Obs1 and Obs2. 
As pre-study test cases, we tried to integrate the HEG and MEG spectra over many different time periods to obtain composite spectra using the \textit{Chandra}'s data analysis system \citep[CIAO version 4.12;][]{2006SPIE.6270E..1VF}. As a result, we concluded that: (1) the MEG composite spectra with a higher SNR are suitable for our current study, while the HEG spectra are too weak to generate time-resolved composite spectra with enough counts for a line fit (on the time scale of several tens of ks); (2) Even for the strong spectral lines, the higher-SNR MEG composite spectra can only reflect the averaged coronal plasma dynamics over a time period of at least several tens of ks.

Therefore, we mainly investigated four time intervals with strong flare activities that we respectively labeled as ``D1", ``D2", ``D3", and ``D4" (see \nfig{fig1}) and two relatively quiescent time intervals flare activities that we labeled as ``reference" ones, using time-resolved MEG composite spectra. 
Despite that the reference interval (60$-$100 ks) in Obs1 includes a small short-duration flare (No.5), we confirmed that it still appears to be a good reference as compared to the flaring intervals D1 and D2 (Section 4.1).
The average X-ray luminosity of these two reference time intervals in the wavelength range of 1.8–26 \AA~ is around 1.3 $\times$ 10$^{28}$ erg s$^{-1}$ and 1.7 $\times$ 10$^{28}$ erg s$^{-1}$, respectively. For each time intervals of our interest, we extracted composite spectra from the MEG spectra and their integration time ranges were manually selected based on our pre-study tests. The time range selection first ensures that each of the analyzed spectral lines has enough counts for a line fit (with a flux of at least $\sim$ 10 count bin$^{-1}$), and then should divide flare events into more different evolution phases from the light curves. According to these selection criteria, the number of composite spectra for each of the analyzed spectral lines could be different (see Section 4). In this integrating process, count errors of composite spectra were recomputed using the default methods provided by CIAO. For bins with sufficiently large counts (N $>20-$30), Gaussian statistics are appropriate, so that the 1-$\sigma$ error is given by $\rm \sqrt{N}$; for bins with lower counts, the Gehrels approximation \citep{1986ApJ...303..336G} to confidence limits for a Poisson distribution was used, so that the 1-$\sigma$ error is given by $\rm1+\sqrt{(N+0.75)}$.


The observed line profiles of these composite spectra can be well characterized by a modified Lorentzian function \citep[see the \textit{Chandra Proposers' Observatory Guide\footnote{https://cxc.cfa.harvard.edu/proposer/POG/html/} and}][]{2004ApJ...617..508T}. This function is described by the relation 
\begin{equation}
F(a, \lambda, \Gamma, \beta) = \frac{a}{ \{1 + [\frac{(\lambda - \lambda_{0})}{\Gamma}]^2\}^{\beta} },
\end{equation}
where $a$ is line intensity, $\lambda_{0}$ is line center, $\beta$ is the exponent, and $\Gamma$ is line width. This modified Lorentzian function will be used for further spectral analysis in Section 3.2 and 3.3.

\subsection{Doppler shift measurements}
We selected three strong and isolated Lyman $\alpha$ doublet lines: O~{\sc{viii}} (18.97 \AA, T$_{peak}$$\sim$3 MK), Mg~{\sc{xii}} (8.42 \AA, T$_{peak}$$\sim$10 MK), and Si~{\sc{xiv}} (6.18 \AA, T$_{peak}$$\sim$16 MK), as well as one strong coronal emission line: Fe~{\sc{xvii}} (15.01~\AA, T$_{peak}$$\sim$6 MK), for Doppler shift measurements. Each of these three Lyman $\alpha$ doublet lines have a known theoretical wavelength difference and an intensity ratio of 2:1 in the optically thin stellar coronae. We applied a two-component Lorentzian function with a fixed wavelength difference, intensity ratio, and line width plus a background to fit these observed Lyman $\alpha$ doublet lines (see Section 4). For the isolated Fe~{\sc{xvii}} line, a Lorentzian function was applied.
With the aid of a Markov chain Monte Carlo (MCMC) analysis method (see Appendix A), the position of each selected line was determined. Considering that the coronal temperature of EV Lac is 3$-$20 MK, Doppler shifts of these selected lines can characterize plasma motions of the cool ($\sim$3 MK), warm (5$-$10 MK), and hot ($\sim$16 MK) components of EV Lac's corona, respectively. Meanwhile, due to the negligible radial velocity of EV Lac  (\speed{1.5}) and the low satellite velocity (\speed{1-2}), the observed Doppler shift of each analyzed spectral line should represent the plasma velocity on the star.
  
\subsection{Plasma density and temperature diagnostics}
The variable line intensities of the He-like triplets of Si~{\sc{xiii}} (6.7~\AA, T$_{peak}$$\sim$10 MK) and O~{\sc{vii}} (22 \AA, $\sim$2 MK) and the line pair of Fe~{\sc{xvii}} (17.05/17.10~\AA, T$_{peak}$$\sim$5 MK) were used to estimate the electron density and temperature variations during the flares on EV Lac. 
The relevant triplets of Si~{\sc{xiii}} and O~{\sc{vii}} are each formed by radiative decays from the upper to the ground state, including the resonance transition ($r$: $1s^2{~}^1S_{0}-1s2p{~}^1P_{1}$), the intercombination transition ($i$: $1s^2{~}^1S_{0}-1s2p{~}^3P_{1,2}$), and the forbidden transition ($f$: $1s^2{~}^1S_{0}-1s2s{~}^3S_{1}$). The flux ratio G = $(f + i)/r$ is mainly sensitive to temperature, while the flux ratio R = $f/i$ is sensitive to electron density because increased electron collisions excite the transition from the $3S_{1}$ to the $3P$ state before the former's radiative decays \citep{1969MNRAS.145..241G,1981ApJ...249..821P}. 
Although other He-like triplets of Ne~{\sc{ix}} and Mg~{\sc{xi}} in \textit{Chandra} X-ray spectra had occasionally been used for coronal plasma diagnostics on stars  \citep[e.g.,][]{2002A&A...394..911N}, here we did not undertake an analysis of them, since they were found to be heavily blended with other lines \citep[e.g.,][]{2002ApJ...580L..73G,2004ApJ...617..508T}. 

For the relevant triplet, line parameters were determined through a three-component Lorentzian function. Assuming each component of the triplet has the same line width and a fixed wavelength difference, this function can be expressed as the following
\begin{equation}
F(\lambda) = F(a_{\rm r}, \lambda_{\rm r,0}, \Gamma, \beta) + F(a_{\rm i}, \lambda_{\rm i,0}, \Gamma, \beta) + F(a_{\rm f}, \lambda_{\rm f,0}, \Gamma, \beta)+B_{\rm 0},
\end{equation} where the subscripts, i.e., $\rm r$, $\rm i$, and $\rm f$, respectively represent the three components of the triplet, and B$_0$ is the background flux. 
Based on the known wavelength differences between different components of the triplet and the definitions of line ratios, this relation can be rewritten as the following
\begin{equation}
F_{\lambda} = F(a_{\rm r}, \lambda_{\rm r,0}, \Gamma, \beta, B_{\rm 0}, \rm R, \rm G), 
\end{equation} 
where the R and G ratios are the density- and temperature-sensitive line ratios of our interest, respectively.
Similarly, line ratios and other parameters of the line pair of Fe~{\sc{xvii}} were also determined through a two-component Lorentzian function.
The MCMC analysis method was also used to constrain and estimate each fitting result (see Appendix A).  Composite spectra at different time intervals and their best fitting results are shown in Section 4. Meanwhile, the Doppler shifts measured from the Si~{\sc{xiii}} triplet and the line pair of Fe~{\sc{xvii}} were also provided as a complement to the Doppler shift measurements in Section 3.3.

\begin{figure}
\plotone{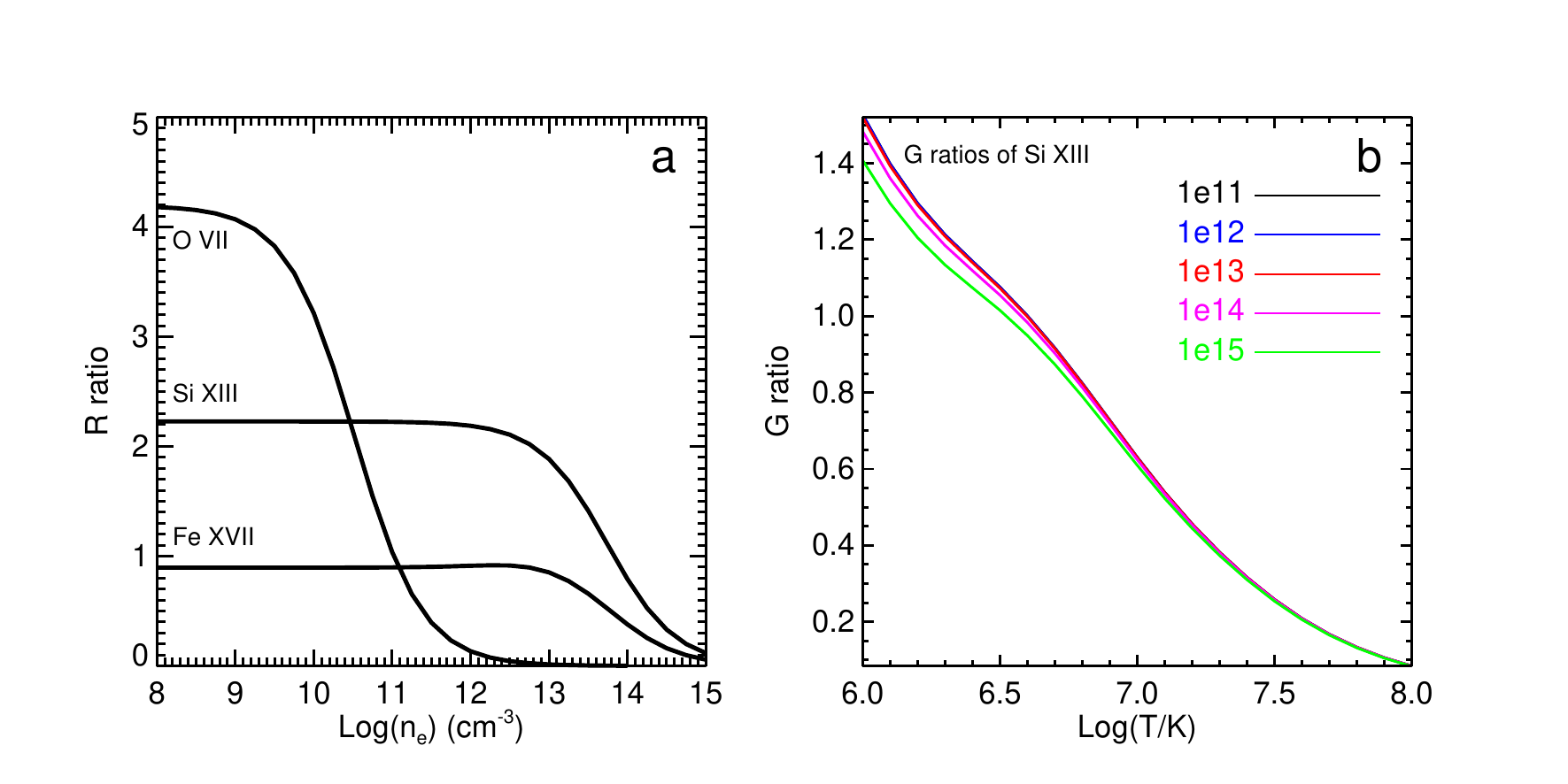}
\caption{Theoretical R- and G-ratio curves of two He-like triplet of Si~{\sc{xiii}} (\AA~) and O~{\sc{vii}}, as well as the line ratio of the line pair of Fe~{\sc{xvii}}, obtained from CHIANTI V.9.0.1. In panel (a), line formation temperatures (10$^7$ K for Si~{\sc{xiii}}, 2$\times10^{6}$ K for O~{\sc{vii}}, and 5$\times10^{6}$ K for Fe~{\sc{xvii}}) were used for the R-ratio calculations. In panel (b), theoretical G ratios of Si~{\sc{xiii}} computed at different density conditions are plotted with different colors. 
\label{fig2}}
\end{figure}

The theoretical curves of R- and G-ratios of the Si~{\sc{xiii}} and O~{\sc{vii}} lines, as well as the line ratio of the Fe~{\sc{xvii}} line pair that we used for the order-of-magnitude estimation of plasma density and temperature are shown in \nfig{fig2}, which were obtained from the CHIANTI atomic database V.9.0.1 \citep{2019ApJS..241...22D}. The line pair of Fe~{\sc{xvii}} are density sensitive when log(n$_e)$ is larger than $10^{13}$ cm$^{-3}$ and their line ratio can well constrain the density as it is below 0.9.
The Si~{\sc{xiii}} triplet are density sensitive when log(n$_e)>10^{12}$ cm$^{-3}$ and the R ratios we measured are mostly near 2.3, so that only an upper limit of density can be given in some time intervals. From \nfig{fig2}(b), we can see that the G ratio of the Si~{\sc{xiii}} triplet only slightly changes with the electron density. So in this work we used the theoretical G-ratio curve computed at the density of 10$^{13}$ cm$^{-3}$ for temperature estimations, because this density is similar to previous density measurement results of the X-ray coronae on many stars from the Si~{\sc{xiii}} triplet \citep{2002A&A...394..911N,2004ApJ...617..508T}. For the much weaker O~{\sc{vii}} triplet, reliable measurements of line ratios  and Doppler shifts in different time intervals are impossible. But its R ratio (around 1.51$\pm$0.28)  measured in the total observed MEG spectra reveals a characteristic coronal density of 5.5$\times$10$^{10}$ cm$^{-3}$, consistent with other previous measurement results \citep[e.g.,][]{2002ApJ...580L..73G,2004ApJ...617..508T,2010A&A...514A..94L}.
Considering the potential model inaccuracies, we would emphasize as much as possible trends in the observed line ratios in this study, rather than the exact density/temperature values derived from the theoretical models.


\section{Results and interpretations}
\subsection{General trends}
In the 200-ks-long observation, we detected significant Doppler shifts during four flaring durations: D1, D2, D3, and D4. Their Doppler velocities in the range of \speed{30$-$110} can not be explained by the spinning motion of coronal structures fixed on the stellar surface, because the velocity of EV Lac is low (\speed{4.5}). The detected Doppler shifts in multi-temperature coronal emission lines result from flare-induced motions of X-ray emitting plasma, since each of their detections is temporally correlated with a flare activity. 
By comparison, Doppler shifts measured in the reference intervals (60$-$100 ks in Obs1 and 0$-$40 ks in Obs2) are almost all below \speed{$\pm$ 20}, which confirms the accuracy and reliability of the wavelength calibration of the \textit{Chandra}/HETGS.

In general, the densities we measured from the line ratios of Si~{\sc{xiii}} are compatible with the low-density limit of 10$^{13}$ cm$^{-3}$ that reported for many stars by \citet{2004ApJ...617..508T}.
Moreover, the variable line ratios of the He-like Si~{\sc{xiii}} triplet in each flaring durations indicate a significant variation of plasma density and temperature with respect to that of the quiescent state. 
Variable ratios of the Fe~{\sc{xvii}} line pair generally indicate an average electron density near $10^{13}$ cm$^{-3}$, but their relevance to the flare activity seems to be more complex. Relatively high densities were found for medium flares, whereas lower densities were found for both the quiescent intervals and strong flares. (see \nfig{fig4}(h) and \nfig{fig6}(h)). Possibly, this is because this line pair have a formation temperature (5 MK) that is close to the background coronal temperature (peaks at log T/K $\sim$6.4) and thus are less sensitive to the high-temperature flaring plasma. So in the following discussion, we will focus on the more illustrative density/temperature diagnostics of high-temperature flaring plasma from the He-like Si~{\sc{xiii}} triplet.

\subsection{Duration D1}

\begin{figure}
\centering
\includegraphics[width=1.\textwidth]{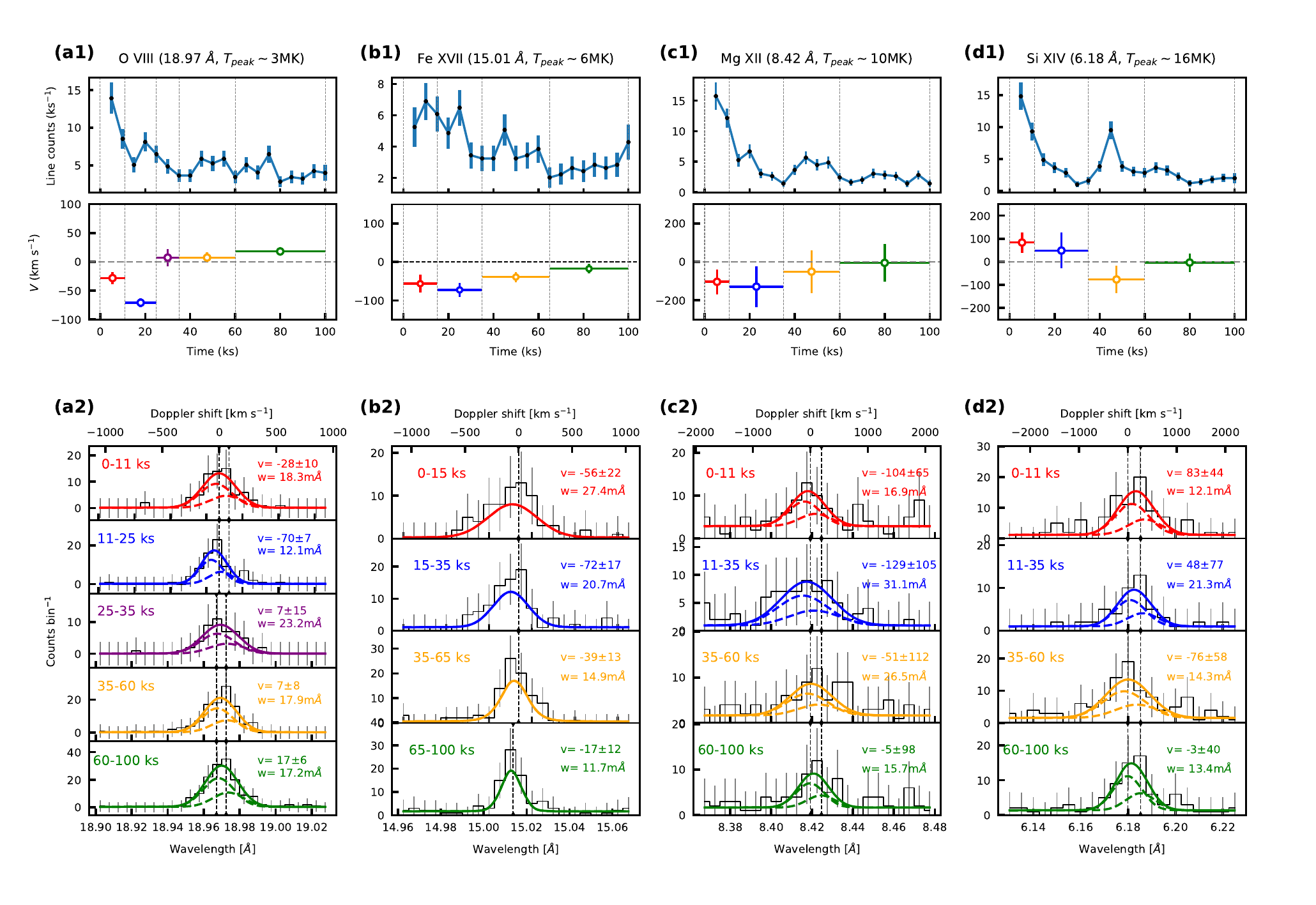}
\caption{Time-resolved spectral line fits of Obs1. Three isolated Ly$\alpha$ doublets from O~{\sc{viii}}, Mg~{\sc{xii}}, Si~{\sc{xiv}} and an isolated Fe~{\sc{xvii}} line averaged over MEG $\pm$1 diffraction orders are analyzed. In all plots the vertical error bars are 1$\sigma$ uncertainties. (a1) Narrow-band light curve of O~{\sc{viii}} (integrated over the wavelength range shown in panel (a2)) and Doppler shifts of O~{\sc{viii}}-emitting plasma derived from time-resolved spectral profiles in panel (a2). For each Doppler-shift estimation, the time interval is indicated by a horizontal bar with different color. (a2) Observed spectral line profiles (black) integrated over different time intervals and their corresponding best fits. Gray vertical dashed lines denote the rest wavelengths of the O~{\sc{viii}} Ly$\alpha$ doublet and the Fe~{\sc{xvii}} line, respectively. Two dashed curves represent the two components of the doublet, solid curve their sum. A different color is used for each time interval, same as that in panel (a1). Similarly, analysis results of the Fe~{\sc{xvii}}, Mg~{\sc{xii}}, and Si~{\sc{xiv}} lines are plotted in (b1-b2), (c1-c2) and (d1-d2), respectively.
\label{fig3}}
\end{figure}

\begin{figure}
\centering
\includegraphics[width=1.\textwidth]{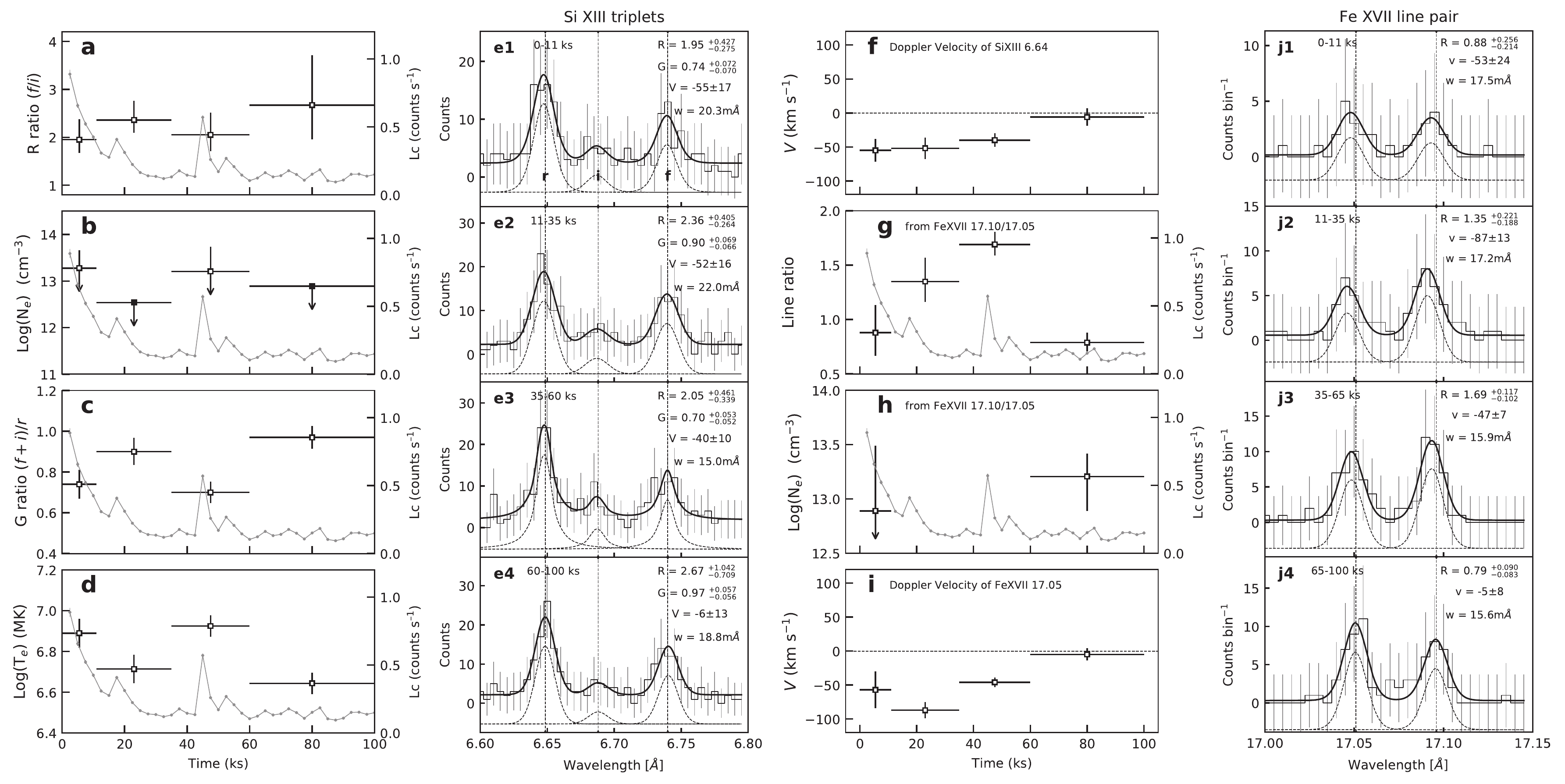}
\caption{Density/temperature diagnostics and Doppler shift measurements using the Si~{\sc{xiii}} triplet and the line pair of Fe~{\sc{xvii}} during Obs1. (a) R ratio, (b) electron density, (c) G ratio, (d) electron temperature, and (f) Doppler shifts derived from spectral fits of the Si~{\sc{xiii}} triplet in different time intervals. (e1-e4) Si~{\sc{xiii}} triple line profiles and (j1-j4) Fe~{\sc{xvii}} line pair profiles in different time intervals and their corresponding line ratios (``R" and/or ``G"), Doppler shifts (``v"), and FWHM (``w") measured from their best-fit results. (g) line ratio, (h) electron density, and (i) Doppler shifts derived from spectral fits of the Fe~{\sc{xvii}} line pair in different time intervals.
The total X-ray light curve (grey thin line, bin size = 2.5 ks) is also plotted for reference. Vertical error bars originate from the measurement errors of the triplet lines, and horizontal bars specify the time intervals of integration. In panels (b) and (h), the available data points with open squares and downward vertical arrows indicate that the lower limits of electron density are not constrained; the data points with filled squares and downward vertical arrows indicate that only the corresponding upper limits of electron density are constrained. The best-fit results for each components of the triplet or the line pair are plotted as dashed lines with a shift along the y-axis, and the black solid curve is their sum. 
\label{fig4}}
\end{figure}

The flaring duration D1 includes the decay phase of a long-duration flare (No.1) and another medium flare (No.2) detected in Obs1 (time range of 0$-$40 ks, see \nfig{fig1}(c) and \nfig{fig3}). The total X-ray energy radiated during D1 is around 7.7$\times$10$^{32}$ erg and its maximum luminosity is about 9.4 $\times$10$^{28}$ erg s$^{-1}$.  During D1, the measured R- and G- ratios of Si~{\sc{xiii}} display an increasing trend, along with the decrease of the X-ray flux (\nfig{fig4} (a) and (c)). In other words, the characteristic coronal density and temperature during F1 are gradually decreasing to a quiescent state (\nfig{fig4} (b) and (d)). The coronal density in flare No.1 was estimated as around 10$^{13.5}$ cm$^{-3}$ (see \nfig{fig4} (b)  {(even with its lower limit unconstrained)}, which is significantly higher than that of the quiescent state in Obs1 (about 10$^{12}$ cm$^{-3}$) and thus indicates a significant density increase in the flaring coronal loops.

The Doppler shift measurements of D1 in \nfig{fig3} and \nfig{fig4} reveal the following behaviors:
in the cool O~{\sc{viii}} line at 18.98~\AA, a blueshift first increases from \speed{28 $\pm$ 10} to \speed{70 $\pm$ 7} in the 0$-$25-ks time interval and then decreases down to a near-zero velocity in the 25$-$35-ks time interval;
obvious blueshifts of \speed{50$-$130} simultaneously appear in many warm lines, including the Mg~{\sc{xii}} line, the Fe~{\sc{xvii}} lines, and the Si~{\sc{xiii}} triplet;
by contrast, in the hot Si~{\sc{xiv}} line, a redshift decreases from \speed{83 $\pm$ 44} to \speed{48 $\pm$ 77}.
This Doppler shift pattern suggests simultaneous hot plasma downflows and cool/warm upflows. We  will discuss the possible origin of such plasma flow pattern in Section 5.

\subsection{Durations D2, D3 and ``gentle evaporation"}
The flaring duration D2 includes two short-duration but relatively strong flares (No.3 and 4) occurring in the 35$-$60-ks time interval of Obs1 (see \nfig{fig1}(c) and \nfig{fig3}). These two flares have peak luminosities of about 6.1 $\times$10$^{28}$ erg s$^{-1}$ and 5.5 $\times$10$^{28}$ erg s$^{-1}$, respectively. The total X-ray flare energy during D2 is about 2.0 $\times$10$^{32}$ erg. 
During these two flares, the measured R- and G-ratio values of Si~{\sc{xiii}} are 2.05$\pm^{0.461}_{0.339}$ and 0.70$\pm^{0.053}_{0.052}$, respectively (\nfig{fig4} (a) and (c)). Compared to the quiescent state in 60$-$100 ks, these line ratios indicate a significant increase of the coronal density up to 10$^{13.4}$ cm$^{-3}$ and a temperature rise (\nfig{fig4} (b) and (d)).  {Note that the density measurement of D2 only gives the best and upper limit values, with the lower limit unconstrained (see \nfig{fig4} (b)).} 

For D2, only one composite spectrum was obtained for each of the analyzed spectral lines. Our Doppler shift measurements of D2 in \nfig{fig3} and \nfig{fig4} demonstrate no obvious Doppler shift at the cool O~{\sc{viii}} line, but simultaneous blueshifts in other warm and hot lines with a velocity of \speed{40$-$80}. 
The Doppler shift pattern suggests the presence of hot and warm coronal upflows without cool downflows. 
Interestingly, the blue shift appears to increase with the line formation temperature, i.e., \speed{40$-$50} in warm lines and \speed{76} in the hot Si~{\sc{xiv}} line, despite the large uncertainty in the latter. Such features are quite similar to the observational characteristics of the gentle chromospheric evaporation observed in some solar flares with a relatively low energy injection rate  \citep[e.g.,][]{1985ApJ...289..425F,2013A&A...557L...5Z,2019ApJ...879...30L}. In this case, the absence of blueshifts in the cool O~{\sc{viii}} can be ascribed to a low injection rate of nonthermal electrons and insufficient dissipation of energy in the lower atmosphere.

The flaring duration D3 occurs in the 40$-$60-ks time interval of Obs2 (see \nfig{fig1}(c) and \nfig{fig5}), which includes two relatively long-duration flares (No. 6 and 8), and a very short-duration flare (No.7). The total X-ray flare energy of D3 is about 2.6 $\times$10$^{32}$ erg and its maximum peak luminosity is about 1.1 $\times$10$^{29}$ erg s$^{-1}$. The plasma flow pattern of D3 and its associated plasma parameter variation measured from the Si~{\sc{xiii}} triplet are also suggestive of gentle evaporation.
Our Doppler shift measurements in \nfig{fig5} and \nfig{fig6} (e2) demonstrate that during D3, blueshifts of several \speed{tens} appear in all cool, warm, and hot spectral lines. Their Doppler velocities also appear to increase with the line formation temperature, i.e., around \speed{40 $\pm$ 10} in the O~{\sc{viii}} line, \speed{43 $\pm$ 4} in the Fe~{\sc{xvii}} line pair, \speed{44 $\pm$ 16} in the Fe~{\sc{xvii}} 15.01~\AA~line, \speed{50 $\pm$ 20} in the Mg~{\sc{xii}} line, \speed{62 $\pm$ 7} in the Si~{\sc{xiii}} triplet, and \speed{69 $\pm$ 36} in the Si~{\sc{xiv}} line. 
Moreover, during D3, the measured R- and G-ratio values of Si~{\sc{xiii}}, as presented in \nfig{fig6}, are 2.03$\pm^{0.191}_{0.165}$ and 1.11$\pm^{0.030}_{0.033}$, respectively. Compared to the quiescent state in 0$-$40 ks, these line ratios also indicate an increase of the coronal density up to 10$^{13.0}$ cm$^{-3}$. However, its temperature appears to show no obvious change (\nfig{fig6} (b) and (d)).

\subsection{Duration D4 and ``explosive evaporation"}
\begin{figure}
\centering
\includegraphics[width=1.\textwidth]{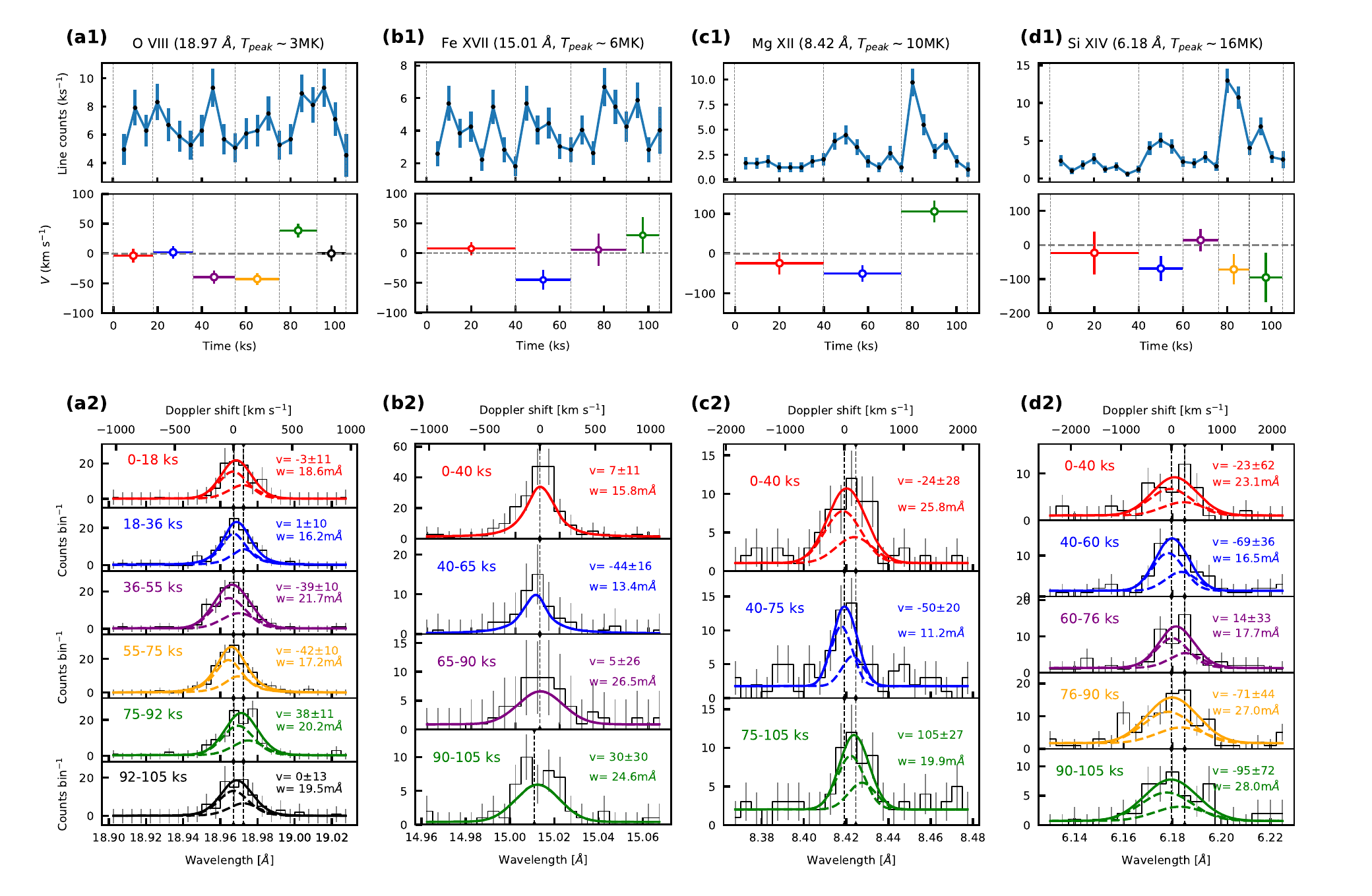}
\caption{Similar to Figure 3 but for Obs2.
\label{fig5}}
\end{figure}

\begin{figure}
\center
\centering
\includegraphics[width=1.\textwidth]{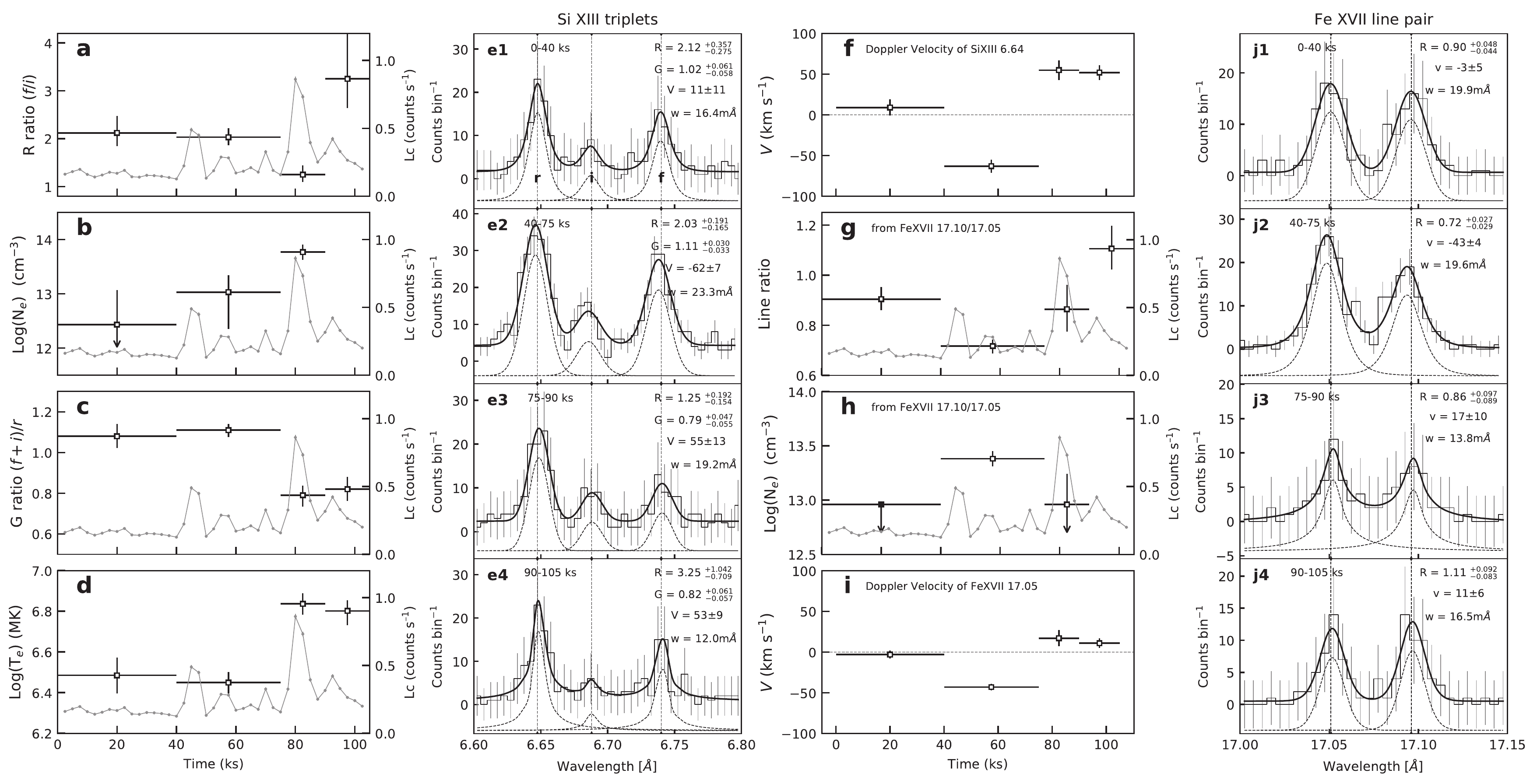}
\caption{Similar to Figure 4 but for Obs2. 
\label{fig6}}
\end{figure}

The flaring duration D4 includes two long-duration flares (No.11 and 12) occurring in 75$-$105 ks of Obs2 (see \nfig{fig1}(f) and \nfig{fig5}). During D2, the peak luminosity of is about 9.1 $\times$10$^{28}$ erg s$^{-1}$ and the total X-ray flare energy is about 6.7 $\times$10$^{32}$ erg. Similarly, the plasma density and temperature diagnosed in D4 from the Si~{\sc{xiii}} triplet also reveal a significant increase compared to the quiescent state (\nfig{fig6}). In particular, the highest plasma density (10$^{13.8\pm0.1}$ cm$^{-3}$) measured in Obs2 is temporally correlated with the flare peak of the event No.11, which is also accompanied with a rise in temperature from the pre-flare level of log T/K = 6.5 to log T/K = 6.8. 

For D4, one composite spectral profile was obtained for the warm Mg~{\sc{xii}} line, and two for the other selected spectral lines. Different from the smaller Doppler shift detected in the pre-flare state (see the time intervals 0$-$18 ks and 18$-$36 ks in \nfig{fig5}), our Doppler shift measurements in \nfig{fig5} and \nfig{fig6} indicate a simultaneous presence of cool/warm coronal downflows and hot coronal upflows in D4. The hot upflows appear as a significant blueshift at the 16-MK Si~{\sc{xiv}} line, which shows a possible velocity increase from \speed{71 $\pm$ 44} in the 76$-$90-ks period to \speed{95 $\pm$ 72} in the 90$-$105-ks period. For the warm downflows, their Doppler velocities are \speed{54 $\pm$ 11} at the Si~{\sc{xii}} triplet and \speed{105 $\pm$ 27} at the Mg~{\sc{xii}} line, respectively. Cooler downflows are simultaneously detected at the O~{\sc{viii}} line and the Fe~{\sc{xvii}} lines, with Doppler velocities in the range of \speed{30$-$45}.

Such a unique plasma flow pattern is highly analogous to that of the explosive chromospheric evaporation process observed in many large solar flares \citep[e.g.,][]{2009ApJ...699..968M,2015ApJ...811..139T,2015ApJ...813...59L}. In this case, the blueshift in the Si~{\sc{xiv}} line can be ascribed to the high-speed evaporating flows at a temperature of $\sim$16 MK, and the redshift detected in the Mg~{\sc{xii}} line and the Si~{\sc{xiii}} triplet most likely results from on-going cooling in post-flare loops. Such downward cooling flows in the 10-MK warm lines likely correspond to the ``warm rain" phenomena, which have been reported in the explosive evaporation of solar flares with a formation temperature higher than the background coronal temperature but lower than that of the flaring loops \citep[e.g.,][]{2003ApJ...586.1417B,2011ApJ...727...98L}.
The redshifts of \speed{30$-$45} detected in the cooler O~{\sc{viii}} line and the Fe~{\sc{xvii}} lines may be caused by cold rains \citep[e.g.,][]{2015ApJ...811..139T,2020PPCF...62a4016A,2021ApJ...910...82L,2022A&A...659A.107C} in post-flare loops and/or chromospheric condensation driven by continuing nonthermal heating. As revealed by many solar flare cases, the appearance of ``warm rain" and possible chromospheric condensation both can be regarded as supportive signatures of explosive injections of non-thermal electrons during flares \citep[e.g.,][]{1984SoPh...93..105I,1985ApJ...289..425F,2003ApJ...586.1417B}.
In particular, the peak characteristic coronal density of D4 is found to rise up to 10$^{13.8}$ cm$^{-3}$, which indicates that a huge amount of heated chromospheric plasma is continuously replenished into the flare loops. Compared to D2 and D3, D4 indeed includes more powerful flare activities with X-ray energy up to 6.7$\times$10$^{32}$ erg, which implies that a higher energy injection rate may be responsible for the observed explosive evaporation process.

\section{Discussion and Conclusion}

The occurrence of hot evaporation flows during stellar flares has long been predicted by the standard flare model, but hot evaporation plasma flows have been hardly detected in coronal emission lines in the past, as mentioned by \citet{2010SSRv..157..211G} and \citet{2019LNP...955.....L}. Alternatively, \citet{2002ApJ...580L..73G} first provide X-ray spectroscopic evidence for chromospheric evaporation during a long-duration flare on Proxima Centauri, through measuring increasing electron densities in the flaring plasma. 
Similarly, the line ratios we measured from the He-like Si~{\sc{xiii}} triplet also indicate a significant increase of coronal density and temperature during four flaring durations (D1, D2, D3, and D4) compared to pre-flare/quiescent states,  {despite that the lower limits of the density estimations are unconstrained during D1 and D2}. In particular, the maximum density we measured appears in D4 and it reaches up to 10$^{13.8}$ cm$^{-3}$, which indicates that a significant amount of hot plasma is explosively replenished into the corona of EV Lac within a few hours. These results conform the scenario of chromospheric evaporation and reveal the possible important role of frequent flare activity in sustaining the million-degree hot coronae above cool stars \citep[i.e.,][]{1991SoPh..133..357H,2014FrASS...1....2P}

Recently, \citet{2019NatAs...3..742A} detected upward/downward motions of hot coronal plasma in the temperature range of 10–25 MK during a long-duration strong flare on a giant star HR 9024, with velocities of \speed{100$-$400}. They claimed that these plasma motions are in agreement with a model of a flaring magnetic tube, in which hot evaporation upflows first appear at the flare onset and downflows then dominate in the decay phase. More importantly, after the flare, they also detected a blueshift of \speed{about 90}, i.e., upflow of cool plasma (about 3 MK), in the O~{\sc{viii}} line, which they ascribed to a possible CME event since it involves only cool plasma. 
In the current work, with time-resolved high-resolution X-ray spectra from \textit{Chandra}/HETGS, we also detected clear evidence for hot coronal evaporation flows during three flaring durations (D2, D3, and D4) on EV Lac using high-temperature coronal emission lines (including the Si~{\sc{xiv}}, Si~{\sc{xiii}},  and Mg~{\sc{xii}} lines). 
Combined with the flare-induced temperature increases, we suggest that the plasma flow patterns observed in D2, D3, and D4 can be explained by flare-induced chromospheric evaporation in the context of the standard solar flare model. Interestingly, despite the relatively large uncertainty in some weak spectral lines, the blue/red shifts detected in D2, D3, and D4 appear to increase with the line formation temperature, which is well consistent with the situations observed in solar cases \citep[i.e.,][]{2008ApJ...680L.157M}.
For D2, we see hot/warm upflows with a Doppler velocity of \speed{several tens} and no obvious cool downflows. For D3, we see cool, warm, and hot upflows with a Doppler velocity of \speed{several tens}. These features are analogous to the gentle evaporation scenario. For D4, we simultaneously detect 16-MK hot upflows with a Doppler velocity of up to \speed{100}, 10-MK warm downflows of \speed{50$-$100}, as well as cooler downflows of \speed{15$-$40}, which is consistent with the scenario of explosive evaporation. The downflows likely result from two alternative scenarios or their superposition: chromospheric condensation and plasma cooling in the flare loops. To the best of our knowledge, this is the first time that high-temperature plasma flows of flare-induced chromospheric evaporation are detected on M dwarfs.

\citet{2019ApJ...875...93C} recently studied the flare-induced plasma flows during two X-class solar flares, through Doppler shift measurements with a Sun-as-a-star EUV spectrometer EUV Variability Experiment on board the \textit{Solar Dynamics Observatory}. This work provides a valuable reference for coronal dynamics of stellar flares. In one of their flare events, we noticed that the hot component of chromospheric evaporation  (around 6.5 MK) is characterized by a strong blueshift of \speed{120$-$200} in the Fe {\sc{xviii}} 94 \AA~line, which rapidly increases in the impulsive phase and lasts for about three hours. Meanwhile, the presence of cold downflows is indicated by a significant redshift of several tens to \speed{170} in the Fe~{\sc{viii}} 131 \AA~line (about 0.37 MK) and lasted for nearly two hours. In general, the coronal upflows/downflows we detected in D2 and D4 show similar lifetimes (on the order of 4$-$7 hours) and Doppler velocities (a time-averaged velocity of several tens to \speed{130}) as in \citet{2019ApJ...875...93C}. This reinforces the conclusion that the upward plasma flows we detected in D2, D3, and D4 result from chromospheric evaporation and that downward plasma flows in D4 are possibly caused by subsequent cooling and/or chromospheric condensation induced by the non-thermal heating. 

Moreover, the Sun-as-a-star spectral observations of \citet{2019ApJ...875...93C} revealed that the temperature at which the Doppler shift transits from blue to red during solar explosive evaporations is close to 1 MK, much higher than that predicted by the chromospheric evaporation model \citep{1985ApJ...289..434F}. While other spatially resolved spectroscopic observations revealed that this reversal can even occurs at temperatures of 2$-$5 MK during some solar flares \citep{2008ApJ...680L.157M,2009ApJ...699..968M,2011ApJ...727...98L}. 
In comparison, here, a much higher reversal temperature, at least up to 10 MK, was observed during our stellar flare events in D4. This higher reversal temperature might be attributed to (1) the different energy deposition rates, heights, or duration driven by magnetic reconnection \citep{2009ApJ...702.1553L} during stellar flares and/or (2) the very different plasma environment (i.e., density and temperature, etc.) in the flaring corona of EV Lac. In addition, such redshifts of hot coronal emission lines might also be expected if thermal conduction effect is strongly suppressed in flare reconnection regions \citep{2015PhPl...22j1206I}.

Compared to D2, D3, and D4, this plasma flow pattern in D1 is more unique and complex (see Section 4.2), because a mega flare (No.1)  and a medium flare (No.2) took place in succession (see \nfig{fig1} (c)). That is to say, the plasma flows and density/temperature variations measured from the composite spectra possibly originate from the superposition of these two flares.
The redshift of the hot Si~{\sc{xiv}} line decreases from \speed{83} to \speed{48} along with the decrease of the X-ray flux. We suggest that it likely results from the plasma emission in the post-flare loops of the mega flare, in which a bulk of hot plasma was trapped and was moving downward at its decay phase. This possibility is also supported by the fact that the narrow-band light curve of the 16-MK Si~{\sc{xiv}} line (\nfig{fig3}(d1)) reveals this mega flare but shows no an obvious signature of the medium flare that peaks at about 20 ks (\nfig{fig1} (c)). 
The cool and warm upflows detected in the time period of 0$-$15 ks might be caused by an on-going chromospheric evaporation at the decay phase of the mega flare, because they are accompanied by a significant increase of plasma density and temperature (as compared to the quiescent period) as measured from the Si~{\sc{xiii}} triplet (in \nfig{fig4} (a-d)). By comparison, the significant cool (3 MK) upflows in the time period of 11$-$25 ks and warm (5$-$10 MK) upflows in the time period of 15$-$35 ks are most likely induced by the medium flare. But they appear to be associated with a decrease of plasma density and temperature at 10 MK (\nfig{fig4} (a-d)). This excludes the possibility of chromospheric evaporation, because in such a case, the 10-MK upflows are expected to be accompanied by an increasing density of the corresponding flaring plasma. Instead, we suggest that this is most likely caused by a plasma ejection event coupled to the flare, i.e., filament/prominence eruption, as reported by \citet{2020MNRAS.499.5047M} on EV Lac and by \citet{2022arXiv220411722X} from Sun-as-a-star observations. In this scenario, the expanding magnetized structure may erupt with cool and warm materials, thus naturally resulting in the simultaneous blueshifts of cool/warm lines and a decreasing plasma density. Note that the upward velocities of these cool/warm erupting materials are far below the escape velocity, thus whether a corresponding stellar CME was successfully launched remains unclear \citep[e.g.,][]{2019NatAs...3..742A,2020MNRAS.499.5047M}.

To summarize, using time-resolved X-ray spectroscopic observations provided by the \textit{Chandra}/HETGS, we detected distinct flare-induced plasma flows in the corona of EV Lac but none of them supports a definite occurrence of stellar CMEs. In the observed spectral profiles of coronal emission lines formed in the temperature range of 3$-$16 MK, these flare-induced upflows/downflows have velocities of several tens to \speed{130}. In most cases, the detection of hot upflows is accompanied by a simultaneous rise in the coronal plasma density and temperature, as inferred from the line ratios of the Si~{\sc{xiii}} triplet. 
Meanwhile, the upflow velocity generally increases with temperature.
We suggest that these flare-induced spectral signatures are X-ray evidence for chromospheric evaporation during flares on EV Lac. In one flare, cool/warm upflows of \speed{50$-$130} were detected, together with a decreasing plasma density. We suggest that this is most likely caused by a stellar filament/prominence eruption coupled to this flare. These results provide new and important clues to understand the coronal plasma dynamics during flares on other stars.

\begin{acknowledgments}
This work was supported by NSFC grants 12103005 and 11825301, and the National Postdoctoral Program for Innovative Talents (BX20200013), and the China Postdoctoral Science Foundations (2020M680201 and 2021M700246). We thank the referee for the helpful comments and constructive suggestions. H.C.C thanks Prof. Junfeng Wang  (王俊峰), Dr. Song Wang (王松), Dr. Jiao Li (李蛟) and Yajie Chen (陈亚杰) for helpful suggestions and discussion.
\end{acknowledgments}

%




\appendix
\section{The MCMC analysis of multi-parameter spectral line fits}
\begin{figure}
\plotone{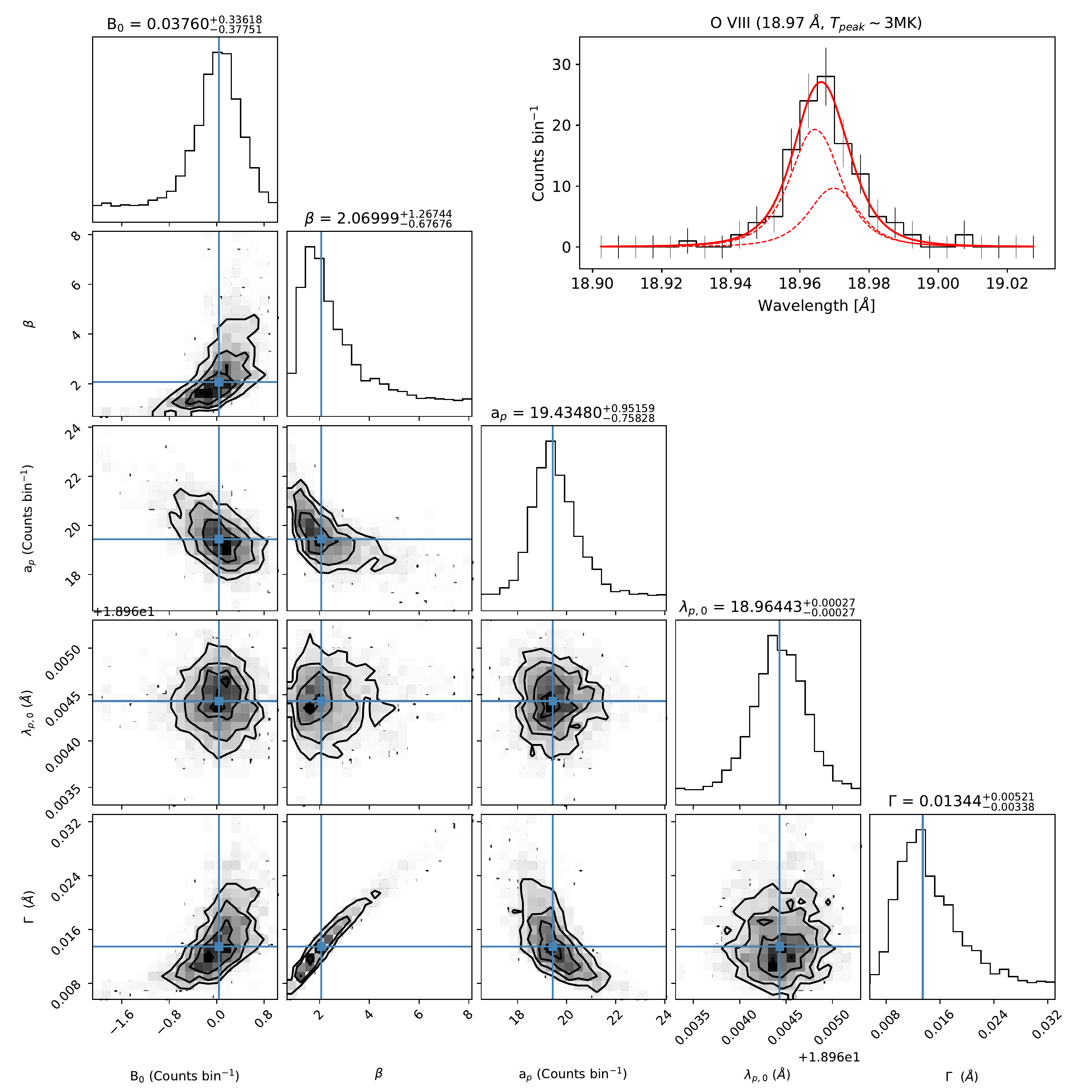}
\caption{MCMC analysis for an example two-component spectral fit of the Ly$\alpha$ doublet used in Figure 3 and 5. Left bottom: Corner plot of the final fit of the isolated O~{\sc{viii}} doublet, which shows posterior distributions of fitting parameters: background flux ($B_0$), the exponent ($\beta$), line intensity of the primary component ($a_{\rm p}$), line center of the primary component ($\lambda_{\rm p,0}$), as well as line width ($\Gamma$). For each parameter, one-sigma uncertainty is reported based on Bayesian statistics.
Right top: The fit result and its observed spectral profile. Two red dashed curves represent the two transitions of the doublet, red solid curve their sum. 
\label{figA1}}
\end{figure}

\begin{figure}
\plotone{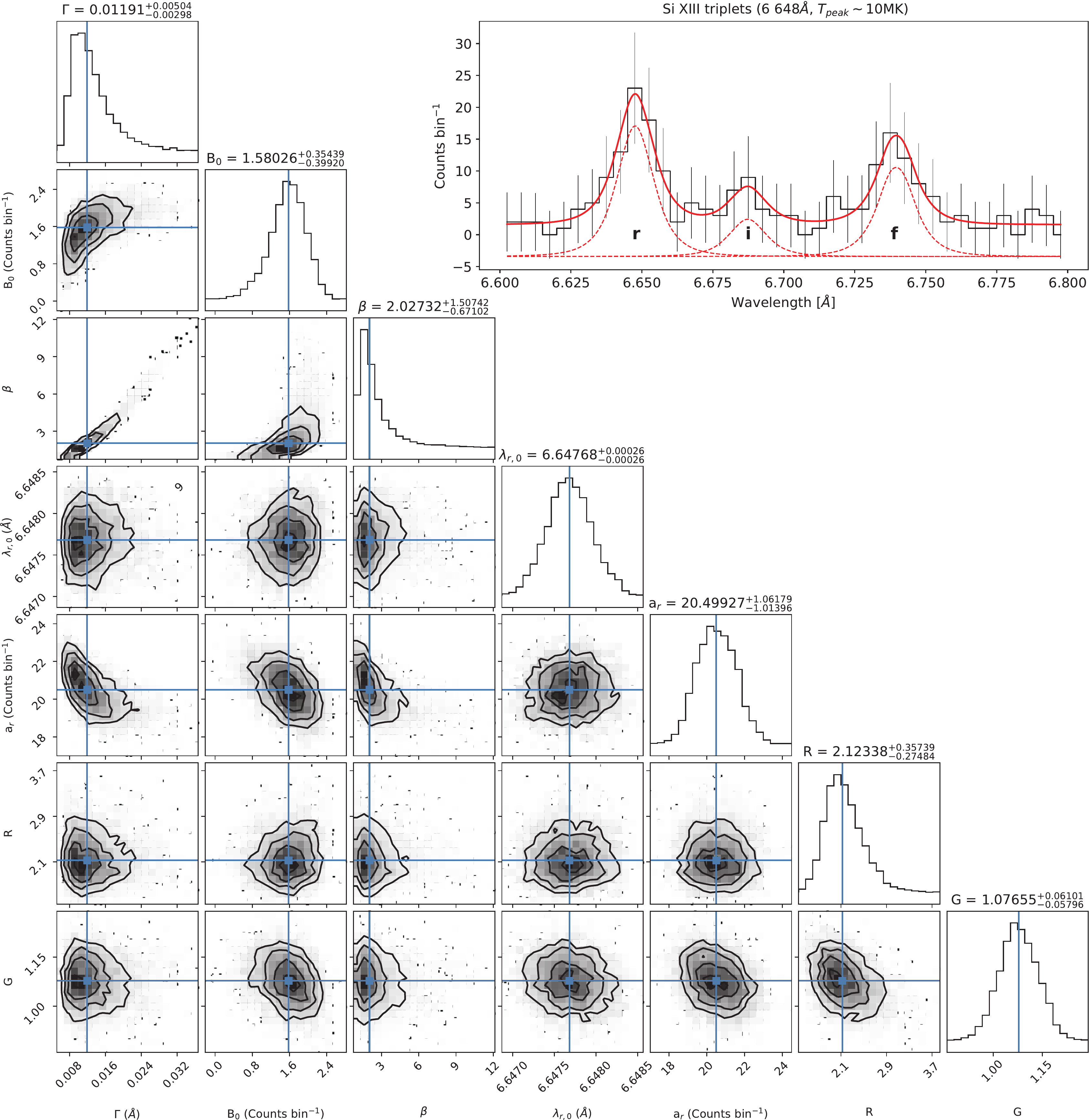}
\caption{MCMC analysis for an example three-component spectral fit of the He-like triplets used in Figure 4 and 6. Left bottom: Corner plot of the three-component spectral fits of the Si~{\sc{xiii}} triplet, which shows posterior distributions of line parameters: width ($\Gamma$), background intensity (B$_{\rm 0}$), exponent ($\beta$), center ($\lambda_{\rm r,0}$), and amplitude ($a_{\rm r}$), as well as R and G ratios of the triplet. Right top: The fit result and its observed spectral profile. For each parameter, one-sigma uncertainty is reported based on Bayesian statistics. The best-fit results of the relevant triplet are plotted as dashed lines with a shift along the y-axis, and the black solid curve is their sum.
\label{figA2}}
\end{figure}

In our Doppler shift measurements and plasma density/temperature diagnostics, a Markov chain Monte Carlo (MCMC) analysis method was used to constrain and evaluate the multi-parameter fitting results. This method can sample the posterior probability distributions of the fit results based on Bayesian statistics and has been extensively used by the astronomical community \citep[e.g.,][]{2019ApJ...875..127L,2020NatAs...4.1140C,2021RAA....21..292W}. The MCMC analysis method we used is an open-source code named \textit{emcee} \citep{2013PASP..125..306F}, which can be found in the Python package LMFIT \citep[see][]{2014zndo.....11813N}.
For each composite spectra obtained at different time intervals, we performed such MCMC analysis. Here, two examples of the MCMC analysis results are shown in the form of so-called ``corner plot" (see \nfig{figA1} and \nfig{figA2}). 
By convention, the non-diagonal and diagonal panels in the corner plot show the two-dimensional projections of the probability distributions between pairs of the fit parameters and the one-dimensional projection of the probability distributions of the fit parameters, respectively. Each fitting parameter is determined with the peak value of probability distributions and its associated one-sigma uncertainty is reported based on Bayesian statistics. Meanwhile, the observed spectral profiles and its best-fitting result are presented at the upper right of the corner plot. We performed the MCMC analysis for two-component spectral fits of the Ly$\alpha$ doublets with 400 steps and 100 walkers, and we discarded the first 100 steps as burn in. For the three-component spectral fits of the He-like triplets, we conducted the MCMC analysis using with 800 steps and 100 walkers, and we discarded the first 200 steps as burn in. As shown in \nfig{figA1} and \nfig{figA2}, all the fitting results in our examples have been well constrained because their probability distributions are all clustered around the maximum likelihood. This thus ensures the robustness of the multi-parameter fitting results we discussed in Section 4. In addition, the full-width-at-half-maximum (FWHM) from each best-fit result is also printed in Figures 3$-$6. According to the \textit{Chandra}'s guide, the average instrumental line width ($\sigma$) for the first-order-diffraction MEG spectra is $\sim$ 9.8 m\AA, which varies with wavelength \citep{2017A&A...607A..14A}. So the FWHM ($\sim2.35\sigma$) for most MEG spectral lines should fall in the range of 16$-$23 m\AA. For the selected spectral lines with a count rate $>$ 15 counts bin$^{-1}$, their observed line widths are in fact roughly compatible with this predicted range and are not significantly broadened during the flare activity. Therefore, velocity dispersion of selected spectral lines along the line of sight is expected to be relatively small.

\bibliography{sample631}{}
\bibliographystyle{aasjournal}

\clearpage

\end{CJK*}
\end{document}